\documentclass[aps,letterpaper,nofootinbib,preprint,showpacs,amsmath,amssymb,pdftex11pt]{revtex4}%
\usepackage{latexsym}%
\usepackage[dvips,dvipdfmx,hypertex]{hyperref}%

\def\cF{{\mathcal F}}
\def\cH{{\mathcal H}}

\def\cW{{\mathcal W}}

\DeclareMathAlphabet{\mathpzc}{OT1}{pzc}{m}{it}

\newcommand{\beq}{\begin{equation}}
\newcommand{\beqn}{\begin{equation}\nonumber}
\newcommand{\eeq}{\end{equation}}
\newcommand{\bea}{\begin{eqnarray}}
\newcommand{\bean}{\begin{eqnarray}\nonumber}
\newcommand{\eea}{\end{eqnarray}}

\begin{document}

\begin{center}
{\bf{\Large Black holes as Gravitational Atoms\footnote{This essay received the Second Award in the 2014 
Gravity Research Foundation Essay Competition.}}}
\bigskip
\bigskip
\bigskip
\bigskip

Cenalo Vaz\\
\bigskip

{\it Department of Physics,}\\
{\it University of Cincinnati,}\\
{\it Cincinnati, Ohio 45221-0011, USA}\\
{email: Cenalo.Vaz@UC.Edu}
\medskip

\end{center}
\bigskip
\bigskip
\bigskip
\bigskip

Recently, Almheiri et.al. argued, via a delicate thought experiment, that it is not consistent to
simultaneosuly require that (a) Hawking radiation is pure, (b) effective field theory is valid
outside a stretched horizon and (c) infalling observers encounter nothing unusual as they cross
the horizon. These are the three fundamental assumptions underlying Black Hole Complementarity
and the authors proposed that the most conservative resolution of the paradox is that (c) is 
false and the infalling observer burns up at the horizon (the horizon acts as a ``firewall''). 
However, the firewall violates the equivalence principle and breaks the CPT invariance of quantum 
gravity. This led Hawking to propose recently that gravitational collapse may not end up producing 
event horizons, although he did not give a mechanism for how this may happen. Here we will support 
Hawking's conclusion in a quantum gravitational model of dust collapse. We will show that continued 
collapse to a singularity can only be achieved by combining two independent and entire solutions 
of the Wheeler-DeWitt equation. We interpret the paradox as simply forbidding such a combination, 
which leads naturally to a picture in which matter condenses on the apparent horizon during 
quantum collapse.
\vskip 2.0in
%%%%%%%%%%%%%%%%%%%%%%%%%%%%%%%%%%%%%%%%%%%%%%%%%%%%%%%%%
%\baselineskip 19pt
	\vfill\eject\

Classical collapse models suggest that a sufficiently massive self-gravitating system will undergo
continued collapse until a singularity forms. In 1975, Hawking \cite{haw75} pointed out that if an event
horizon forms and if effective field theory is valid away from a stretched horizon, then
radiation from the black hole is produced in a mixed state from the point of view of the observer who
remains outside the black hole provided that the freely falling observer detects nothing unusual (``no drama'')
while crossing the horizon. Under these conditions information is lost if the black
hole evaporates completely, which violates unitarity and led Hawking to propose that quantum mechanics 
should be modified \cite{haw76} (he has since changed his mind). To preserve unitarity in quantum 
mechanics one of two possibilities must be true: (a) Hawking radiation is 
in fact pure or (b) the evaporation leaves behind a long lived remnant, which preserves all the information
that collapsed into the black hole. However, if quantum gravity is CPT 
invariant then remnants are ruled out and only the first of the two options above remains viable. In 1993, 
building on the work of 't Hooft \cite{thft90} and Preskill \cite{pres92}, Susskind et. al. \cite{sus93} proposed 
that unitarity could be preserved if information is {\it both} emitted at the horizon {\it and} passes 
through the horizon so that an observer outside would see it in the Hawking radiation and an observer who 
flies into the black hole would see it inside. No single observer would be able to confirm both pictures: 
one simply cannot say where the information in the Hilbert space is located, so quantum 
mechanics is saved at the cost of locality. This is the principle of Black Hole Complementarity

Recently, Almheiri et. al. (AMPS) \cite{alm12} suggested that the three assumptions of Black Hole Complementarity
{\it viz.,} (a) unitarity of Hawking evaporation, (b) validity of effective field theory outside a
stretched horizon and (c) ``no drama'' at the horizon for a freely falling observer are not self-consistent.
Briefly, their argument can be stated as follows. Consider a very large black hole so that a freely falling
observer crossing the horizon sees an effectively flat spacetime (on scales much smaller than the horizon
length). From the point of view of an observer who stays outside the horizon, the purity of the Hawking radiation 
implies that late time photons are maximally entangled with some subset of the early radiation. However, 
these late photons when propagated back from infinity to the near horizon region using effective field theory must 
be maximally entangled with modes inside the horizon from the point of view of the freely falling observer (this 
is simply a property of the Minkowski vacuum, appropriate to a freely falling observer). This is not permitted by 
the strong additivity of entanglement entropy. Assuming then
that effective field theory is valid and that Hawking radiation is pure, the paradox can only be avoided if the 
backward propagated photon is not entangled with a mode behind the horizon. But this would lead to a divergent stress
tensor near the horizon, so AMPS concluded that the freely falling observer would burn up before she
could cross it. This is the ``firewall''.

Considerable interest has surrounded the proposed firewall \cite{fiw}, all of it assuming that continued collapse 
will occur, leading to black holes with event horizons. But Hawking has recently raised several objections to the 
firewall and suggested that the correct resolution of the AMPS paradox is that event horizons do not form, only 
apparent horizons form \cite{haw14}. Radiation from the black hole is then deterministic, but chaotic. In this essay, 
we will justify this proposal in the context of the exact quantum collapse of spherically symmetric dust, showing 
that in fact continued collapse can only be achieved by artificially combining two independent solutions of the 
Wheeler-DeWitt equation, each of which covers the entire spacetime.

The classical spherical collapse of inhomogeneous dust in AdS of dimension $d=n+2$ is described by the
LeMaitre-Tolman-Bondi (LTB) family of metrics \cite{ltb}. The models may be expressed in canonical form
after a series of simplifying canonical transformations and after absorbing the surface terms
\cite{kuc94,brkk95,va1,va2}. They are then described in the phase space consisting of
the dust proper time, $\tau(t,r)$, the area radius, $R(t,r)$, the mass density, $\Gamma(r)$, and
their conjugate momenta, $P_\tau(t,r)$, $P_R(t,r)$ and $P_\Gamma(t,r)$ respectively, by two
constraints,
\bea
&&\cH_r = \tau' P_\tau + R' P_R - \Gamma P_\Gamma' \approx 0\cr
&&\cH = P_\tau^2 + \cF P_R^2 -\frac{\Gamma^2}{\cF} \approx 0,
\eea
where 
\beq
\cF~ \stackrel{\text{def}}{=}~ 1-\frac{F}{R^{n-1}} + \frac{2 R^2}{n(n+1)l^2}.
\eeq
with $\Lambda=-l^{-2}$ representing the cosmological constant and $F(r)$ the mass function. The condition $\cF = 0$ 
determines the physical radius of the apparent horizon and is an essential singularity of the wave equation. Dirac 
quantization of the constraints leads to a Wheeler-DeWitt equation which, for a smooth dust distribution, 
can be regularized on a lattice. Each point (labeled by ``$i$'' below) on the lattice will then represent a 
collapsing dust shell. Assuming that the wave-functional is factorizable and taking $\sigma$ to be the lattice spacing, 
it can quite generally be written as
\beq
\Psi[\tau,R,\Gamma] = \lim_{\sigma\rightarrow 0} \prod_i \psi_i(\tau_i,R_i,F_i) = \exp\left[-\frac i\hbar 
\int dr \Gamma(r) \cW(\tau(r),R(r),F(r))\right],
\label{wfnal}
\eeq
where each $\psi_i$ resides on the lattice point $i$ and can be thought of as a shell wave function. The wave functional
automatically obeys the momentum constraint provided that $\cW(\tau,R,F)$ has no explicit $r-$dependence. Independence 
of the wave functional on the lattice spacing implies that the lattice wave functions must satisfy three equations 
\cite{va2,va3}, one of which is the Hamilton--Jacobi equation, which was used to describe the Hawking radiation in 
\cite{va2}. The other two equations together uniquely fix the Hilbert space measure and the factor ordering. For the shell 
wave functions, $\psi_i$, one finds the exact positive energy solutions
\beq
\psi_i = e^{\omega_i b_i} \times \exp\left\{-\frac{i\omega_i}\hbar \left[a_i \tau_i \pm \int^{R_i}
dR_i \frac{\sqrt{1-a_i^2\cF_i}}{\cF_i}\right]\right\},
\eeq
where $a_i = 1/\sqrt{1+2E_i}$ is related to the energy function, $\omega_i = \sigma \Gamma_i/2$ and
the factor $e^{\omega_i b_i}$ is a normalization. 

These wave-functions are well defined everywhere except at the apparent horizon, where there is an essential
singularity. In order to match interior to exterior solutions, we deform the integration path in the
complex $R_i$-plane so as to go around the essential singularity at $\cF_i = 0$ \cite{va4,va5}. This is similar 
to the quasi-classical tunneling approach employed in various semi-classical analyses \cite{pw00} (the deformed 
path does not correspond to the trajectory of any classical particle). The direction of the deformation is chosen 
so that positive energy solutions decay. One finds the following solution representing collapse with support 
everywhere in spacetime \cite{va5}: 
\beq
\psi_i^{(1)}(\tau_i,R_i,F_i) = \left\{\begin{matrix}
e^{\omega_ib_i} \times \exp\left\{-\frac{i\omega_i}\hbar \left[a_i \tau_i + \int^{R_i} dR_i
\frac{\sqrt{1-a_i^2\cF_i}}{\cF_i}\right]\right\} & \cF_i>0\cr
e^{-\frac{\pi\omega_i}{\hbar{g_{i,h}}}}\times e^{\omega_ib_i} \times \exp\left\{-\frac{i\omega_i}\hbar
\left[a_i \tau_i + \int^{R_i} dR_i \frac{\sqrt{1-a_i^2\cF_i}}{\cF_i}\right]\right\} & \cF_i < 0
\end{matrix}\right.
\label{wfn1}
\eeq
where $g_{i,h}$ is the surface gravity at the apparent horizon. It represents dust shells condensing to the apparent 
horizon on both sides of it but the interior, outgoing wave appears with a relative probability of $e^{-2\pi\omega_i/
\hbar g_{i,h}}$, which is the Boltzmann factor for a shell at the ``Hawking'' temperature, $T_{i,H} = \hbar g_{i,h}/2\pi 
k_B$. 

Another independent solution exists, with support everywhere and the same deformation of the integration path, 
\beq
\psi_i^{(2)}(\tau_i,R_i,F_i) = \left\{\begin{matrix}
e^{-\frac{\pi\omega_i}{\hbar{g_{i,h}}}}\times e^{\omega_ib_i} \times \exp\left\{-\frac{i\omega_i}\hbar
\left[a_i \tau_i - \int^{R_i} dR_i \frac{\sqrt{1-a_i^2\cF_i}}{\cF_i}\right]\right\} & \cF_i > 0 \cr
e^{\omega_ib_i} \times \exp\left\{-\frac{i\omega_i}\hbar \left[a_i \tau_i - \int^{R_i} dR_i
\frac{\sqrt{1-a_i^2\cF_i}}{\cF_i}\right]\right\} & \cF_i < 0
\end{matrix}\right.
\label{wfn2}
\eeq
Here, dust shells move away from the apparent horizon on either side of it but this time the exterior, outgoing wave 
is suppressed by the Boltzmann factor at the Hawking temperature for the shell.

In principle we may take the general solution representing the collapsing dust ball to be a linear combination of the 
two solutions \eqref{wfn1} and \eqref{wfn2},\footnote{Strictly speaking, one superposes the wave functionals constructed 
separately out of $\psi_i^{(1)}$ and $\psi_i^{(2)}$ according to \eqref{wfnal}. Then one finds a gauge invariant 
relative probability for the exterior, outgoing wave-functional of $|\mathcal A|^2e^{-S}$, where $\mathcal A = \prod_i A_i$ 
and $S$ is the Bekenstein-Hawking entropy of the AdS black hole \cite{va5}.}  
\beq
\psi_i = \psi_i^{(1)} + A_i\psi_i^{(2)}.
\eeq
However, there is nothing within the theory that suggests a value for $A_i$ and further input is needed to determine 
these amplitudes. Note that if $0<|A_i|\leq 1$, the dust will 
ultimately pass through the apparent horizon in a continued collapse on its way to a central singularity and an event 
horizon will form. This process will be accompanied by thermal radiation in the exterior. We therefore see that the AMPS 
paradox provides the required additional input. To avoid a firewall, $|A_i|$ must vanish and therefore \eqref{wfn1} 
{\it alone} is a complete description of the quantum collapse. But this solution says that each shell will condense to 
the apparent horizon and will not undergo further collapse; there is no tunneling into the exterior and no firewall. 
As each shell converges to the apparent horizon, a ``dark star'' forms. The density profile of such a dark star will 
depend on the initial data, but we can expect that it will attain very high densities in the central regions. Even so, 
provided that the initial data respect cosmic censorship, no central singularity can form.

Thus, in view of the AMPS paradox, an entirely new picture of the black hole has emerged. Instead of a spacetime singularity 
covered by an event horizon we will have an  essentially quantum object, an extremely compact dark star, which is held up 
not by any degeneracy pressure but by quantum gravity just as ordinary atoms are sustained by quantum mechanics. Astronomical 
observations \cite{nm08,bln09} indicate that astrophysical black holes possess dark surfaces and this is consistent with the 
picture we have just described.

\end{document}